# Airplane Type Identification Based on Mask RCNN; An Approach to Reduce Airport Traffic Congestion

W.T AL-SHAIBANI
Department of
*Electrical and Electronics Engineering*
Istanbul Technical University
Istanbul, Turkey
tawfiqwaleed@gmail.com

Mustafa HELVACI
Institute of Informatics,
Satellite Communication, and Remote Sensing Program
Istanbul Technical University
Istanbul, Turkey
helvacim@itu.edu.tr

Ibraheem SHAYEA
Department of
*Telecommunication and electronics Engineering*
Istanbul Technical University
Istanbul, Turkey
ibr.shayea@gmail.com

Azızul AZIZAN
Advanced Informatics Department, Razak Faculty of Technology and Informatics
Universiti Teknologi Malaysia
Kuala Lumpur, Malaysia
azizulazizan@utm.my

*Abstract*— One of the most difficult jobs in remote sensing is dealing with traffic bottlenecks at airports. This fact has been confirmed by several studies attempting to resolve this issue. Among a wide range of approaches employed, the most successful methods have been based on airplane object recognition using satellite images as datasets for deep learning models. Airplane object identification is not a viable method for resolving traffic congestion. There are several types of airplanes at the airport, each with its own set of requirements and specifications. Utilising satellite pictures will require the use of complex equipment, which is a financial burden. In this work, a universal, low-cost and efficient solution for airport traffic congestion is proposed. Drone-captured aerial pictures are used to train and assess a Mask Region Convolution Neural Network model. This model can detect the presence of any aircraft in a photograph and pinpoint its location. It also includes mask estimations to properly identify each detected airplane type based on the estimated surface area and cabin length, which are crucial variables in distinguishing planes. The study is conducted using Microsoft's Common Object in Context (COCO) metrics, average accuracies and a confusion matrix, all of which demonstrate the approach's potential in providing valuable aid for dealing with traffic congestion at airports.

## I. INTRODUCTION

Owing to the dramatic development of technology, many conveniences are now offered throughout numerous fields to support modern human life. One such convenience would be travelling by plane, which is the preferred method of travel for most people. The increasing number of users has forced airlines to increase their productivity. In consequence, many different types of airplanes are used such as commercial aircrafts, private airplanes and shipping airliners. Each aircraft type requires several new routes. Since there has been a vast increase in the types of airplanes used, airports now face the tremendous pressure of controlling massive amounts of flights. The addition of infrastructure can be considered as a temporary solution, yet it remains insufficient in future since it would create an economic impact. Not all airlines throughout the world can increase their airport size due to several limitations concerning geographical location. Unfortunately, airport ground traffic control experiences difficulties when the limited number of flights per day increases, leading to traffic jams. Effective technology is needed to support airport traffic control and address the issue of airport traffic jams. One method employed is airplane detection from satellite images using the deep learning (DL) approach. The dataset is collected from public satellite sources.

Object detection (OD) is a method that automatically determines the existence of objects. It is independent of any human assistance. This field has been widely used for various sectors. The military and defence fields were the first to employ this method before sharing it with other research areas of modern technology. Currently, this approach overlooks tracking, delivery and security issues. OD possesses a key feature which has made it the most popular in terms of remote sensing: it can perform its function of extracting data, analysing and deciding the differentiation between objects without physically touching the objects. The basic idea is to compare data with a threshold so that it can determine the object's existence or occurring actions in case the limit has been exceeded. A clear example of this function would be the detection process that occurs in the human visual system. This process looks for signatures that differentiates objects from their background. The difference between the background and object must be bigger than the threshold in order to be detectable. The current methodology used is nothing like past techniques. The sliding window scanning method is only useful for low-level vision components within a view. Measuring the similarities between shapes, textures, edges, etc., can be used to detect these components. A clear example of current technology is Facebook's automatic tagging; it can recognise and identify faces directly from images. Machine learning (ML) offers a significant leap in the OD field due to the different techniques offered such as statistical appearances model, wavelet feature and gradient-based representation.

The history of object detection can be divided into two sections: classic detection and modern detection. The classic detection relies on handcrafted-extracted features used for approximately all object detection algorithms such as Viola-Jones Detectors [1], Histogram of Oriented Gradients (HOG) [2] and Deformable Part-based Model (DPM) [3]. Viola achieved an excellent accomplishment that allows real-time detection of human faces. It was considered to be significantly faster than other methods of its time. HOG is an enhancement of the scale-invariant feature transform (SIFT) [4], while DPM is an extension of the HOG detector. Modern detection techniques extensively use deep learning due to the limitations in the extraction process of handcraft features. These developments have led to the creation of the convolution neural network (CNN) in object detection. CNN can learn and conduct a robust feature extraction. Deep learning models consist of multiple stages. Some are composed of two stages such as the Region Convolution Neural Network (RCNN) and Feature Pyramid Network (FPN). Others consist of one stage such as the Single Shot Multi-Box Detector (SSD) and You Only Look Once (YOLO).

The RCNN was launched by Ross Girshick to avoid selecting a large number of regions [5]. The function of selective research is to limit the number of extracted regions to 2000 region proposals. These region proposals are then forwarded to a classifier after they are appropriately resized for the neural network since a fully connected neural network does not accept different sizes. Instead of classifying a huge number of regions, only 2000 regions are created and forwarded to the classifier. A fast RCNN was later released by the same author. This approach is similar to RCNN except that it feeds the input image directly to the CNN instead of extracting a region of the proposal then forwarding it. Fast RCNN forwards the input image directly to the classifier to generate a convolution feature map that is used to identify the region of proposals. The significant contribution of this methodology is that it introduces the region of interest pooling layer [6]. Subsequently, a faster RCNN was later released and is now considered to be the first real-time detection approach. The greatest contribution of faster RCNN is the introduction of the Region Proposal Network (RPN) [7].

The rapid growth and development of OD techniques have led to promising results in several areas that apply OD in their applications, making this research field highly interesting. DL has changed the way we see technology, particularly with the massive amounts of abilities embedded with computational power. Computational power allows machines to recognise and identify objects in real time, leading to the creation of numerous applications that affect human life. Generally, DL is included in several systems such as in tools, libraries, universities, government units and research equipment. Complicated functions, such as remote sensing tasks, can now be addressed which have led to significant improvements. Currently, remote sensing applications are more reliable depending on object detection, especially for deep learning-based object detection. There are many DL-based applications that have been recently presented such as self-driving cars, healthcare applications, voice search, the addition of sounds to silent movies, face recognition, automatic translation, automatic text generation, speech recognition, human detection, image recognition, automatic image caption, change detection, automatic colorization, shape detection and airplane detection.

OD is continuously developing every day. Many applications in real life rely on object detection due to its achievements and great state-of-the-art results. Airplane detection is one of the most important research which currently requires further development. It is still considered as an area of open research. In previous years, airplane detection was exclusively for defence and military purposes. However, it is now applied in different fields, encompassing both military and civil areas. Airplanes are used frequently, mostly for traveling and shipping which adds a significant load on traffic ground control. Since traffic ground control cannot efficiently deal with the large number of daily flights, this creates airport traffic jams. Usually, airports have their own schedule for daily flights, but after a long period, this schedule becomes overcrowded with flights. Not all airports are capable of sufficiently controlling such traffic. In future, even small airports will have crowded schedules due to lower flight cost, as noticed in previous years. This is why some countries are developing larger and newer airports to satisfy the huge demand for flights. Many researchers contributed to solving this issue using different approaches. The fastest and most intelligent techniques are those that had implemented deep learning in their approaches. There is a wide range of airplane detection and deep learning-based research using satellite images as datasets. However, this paper uses drone images instead of satellite to feed the system with the dataset. Aerial images taken by drones are typical, simple and straightforward during processing. They also have higher resolution since the elevation is too small compared to satellites. Several options are available to enhance images with the use of various processes such as capturing the moving part, infrared bodies or night mode. Overall, aerial images taken by drones are lower in cost, require less time and have less complicated processing than satellite-based approaches for airplane detection. Figure 1 presents the airplane detection process. First, images are acquired using a remote sensing tool such as a drone or satellite. Next, the acquired images are pre-processed to prepare them for the detection process. The detection process is performed to identify airplane objects from aerial images. Using drone images will reduce the time consumed in performing airplane detection tasks. However, it is illegal to fly drones over airports due to security reasons. Any drone that flies over an airport will be neutralised. To utilise drone images, drones must be placed in specific coordinated points and only used by the airport itself. The employed drones will not fly on a horizontal plane over the airport ground area but at a vertical axis, moving up and down for a specified duration as defined by the airport. The main goal is to acquire aerial images within airports. This can be accomplished by placing 3D cameras on top of several fixed towers. With the help of 3D modelling, the cameras can be used to feed the system with aerial images. The suggested approach provides a useful, general and low-cost solution for airplane detection at airports, as explained in the evaluation section. This approach accomplishes the goal of detecting airplane objects at airports, however, the main issue has not been fully solved. The main issue is addressing airport traffic jams, therefore, knowing the type of detected airplane is crucial. Identifying the airplane type provides useful information regarding the technical and physical needs of each airplane at the airport. Servicing each airplane will become well organised and time efficient. This paper focuses on identifying the type of airplane based on the results of airplane detection. There are two ways to differentiate airplane types: the surface area-based approach and the length-based approach. The surface area-based approach relies on the surface area of the detected airplane which can be found by using the ground resolution with the total number of covered pixels in the detected area. The length-based approach determines the length of any detected plane by measuring the distance between any two farthest points in the detected airplane area.

The rest of this paper is organised as follows: Section II provides the related works. Section III explains the methodology and practical steps. Section IV discusses the results of the evaluation process. Finally, Section V concludes the entire paper.

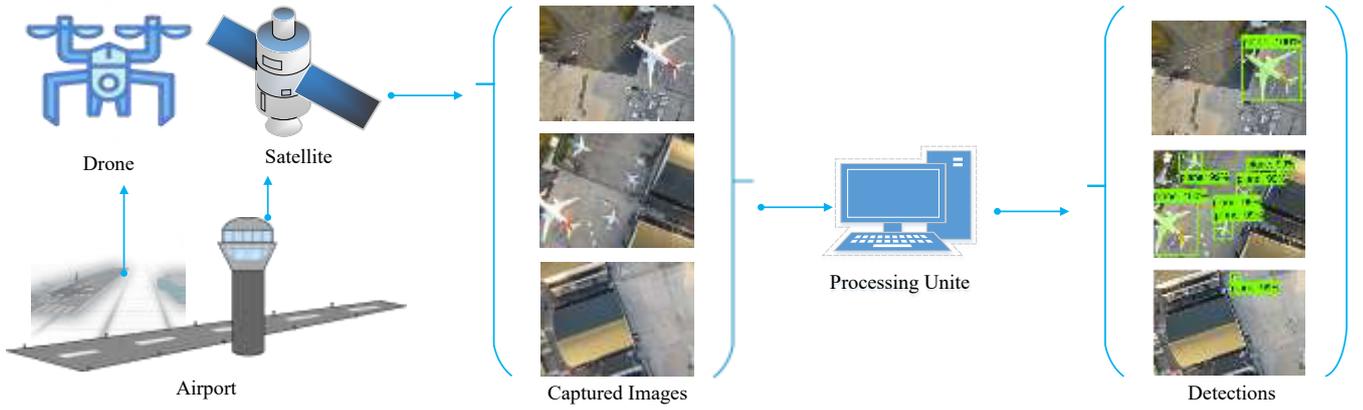

**Figure 1**: Airplane detection process

## II. RELATED WORKS

This section highlights several interesting research previously accomplished on object detection. In [8, 9], the authors used an enhanced method to detect different objects such as boats or airplanes. The training dataset was collected as parts from large images, then an augmentation process was applied using rotations and random saturation scaling. In [10], the author used high-resolution images with many labelled cars from different areas to detect cars using a deep learning model. In [11], the author presented a circle detection approach to detect oil tanks in satellite images. Faster RCNN was firstly used to extract the region of the oil tank object, then an improved co-segmentation with saliency co-fusion was applied to identify bright oil tanks. In [12], the authors introduced edge-boxes to obtain the edge information, and CNN for the feature extractor. CNN and edge-boxes were used to conduct target classification of aircraft objects or non-aircraft objects. A transferred deep model based on AlexNet CNN has being introduced by Zhou et al. [13]. In [14], a Single Shot Detector (SSD), faster RCNN and You Only Look Once (YOLO) were used on satellite images for airplane detection. They divided the collected data into training, evaluation and testing datasets, then created the training process for the proposed models. In [15], the authors introduced a fast region-based convolutional neural network (R-CNN) method for ship detection from high-resolution remote sensing imagery. In [16], the authors suggested that using CNN with a limited number of labelled examples may be problematic since it can lead to reaching the overfitting limit. Thus, they proposed a method that enables the pre-trained CNN to tackle an entirely different type of classification problem called the ImageNet challenge. Their proposed solution was in two stages, and they managed to deal with the limited-data problem. Their methodology includes a novel feature fusion algorithm that tackles large data dimensionality. In [17], the authors introduced a new two-stage framework method. The first stage is a pre-trained CNN designed for tracking an entirely different classification problem then exploiting it to extract an initial set of representations. The second stage is transferring the driven representations into a supervised CNN classifier. In [18], the authors established a land use classification in remote sensing images using CNN. In [19], the authors analysed texture methods applied to remote sensing classification. They created a database with high resolution in both satellite and area images. The texture classes were defined in urban applications. It was crucial to determine the type of vegetation in the first application. Overall, they conducted a comparative study on texture feature extraction for classifying remote sensing data using wavelet decomposition. In [20], the authors classified satellite images with regularised AdaBoosting of RBF neural networks. In [21], the authors introduced a sparse coding method for satellite scene classification. They also presented local ternary pattern histogram Fourier features and combined a set of divers with complementary features to enhance performance. They were able to accomplish high-resolution satellite scene classification using a sparse coding-based multiple feature combination.

## III. SYSTEM AND SIMULATION MODEL

This section introduces the dataset acquisition process. The dataset consists of aerial images taken by two eBee Classic drones flying simultaneously. The targeted area was Le Bourget Airport in Paris [22]. The aerial images have a high resolution of 3.14 cm/px and size dimension of 4608 x 3456 px. The collected dataset was divided into two categories according to size dimensions. The low-resolution category has a dimensional size of 369 x 259 px. The smaller the size, the faster the processing. The high resolution has a dimension size of 922 x 864 px. Table 1 presents both categories in this section.

TABLE I. DATASET CLASSIFICATION

| Dataset Categorisations | | Number of Images |
|---|---|---|
| General images with high-resolution | | 557 |
| Images with low-resolution | General | 557 |
| | With an airplane object | 94 |

To enhance deep learning, data augmentation processing is used. These augmentations vary between noise addition, rotation and cropping. The purpose of this process is to increase the number of images, as illustrated in Table 2

TABLE II. DATASET CLASSIFICATION

| Dataset Categorisations | | Number of Images |
|---|---|---|
| General images with high-resolution | | 1114 |
| Images with low-resolution | General | 1114 |
| | With an airplane object | 689 |

TABLE III. DATASET CLASSIFICATION

| | | Model 2 | Model 3 | | Model 4 | Model 5 | Model 6 | Model 7 | Model 8 |
|---|---|---|---|---|---|---|---|---|---|
| Dataset | | Low resolution with airplane object | Low resolution with airplane object | | Low resolution with airplane object | Low resolution with airplane object | Low resolution with airplane object | High resolution | Low resolution |
| Additional changes in configurations | Image Sizes | 230 X 350 | First_stage_features_stide | 8 | Manual Step Learning Rate= 0.002 Schedual=0,0002-0.00002 | First_stage_max_proposal | Manual_Step_Learning_Rate= 0.00002 Schedual= 0.000002-0.0000002 | - | - |
| | | | Hight_stride | 8 | | 100 | | | |
| | | | Width_stride | 8 | | Learning rate | | | |

*A. Training*

The training mode allows a computing system to make the correct decision. In this paper, the aim is to determine the existence of the airplane object as well as its location inside the image. The mask RCNN model has been applied with different configurations [23]. The aim of this study is to achieve the highest detection performance based on tailored configurations, illustrated as different models. A useful platform called TensorFlow helped in conducting the experimental part of this paper. TensorFlow is a deep learning platform that provides pre-trained modes to initialise the neural network with efficient time usage [24]. As a rule of thumb, the optimal values of hyperparameters are variable. It is not easy to find a typical set of values; they require several different iterations. These values are likely to be different based on changes in the dataset used. Due to these reasons, different changes have been made to the base configuration taken from TensorFlow. Among these iterations, seven models had considerable training responses in addition to the base model. Table 3 illustrates the training models with different configurations.

*B. Airplane Type Identification*

The advantage of estimated masks for each detected airplane object with the ground resolution offered by the dataset has made it possible to calculate the surface area. This paper introduces a direct method of calculating the surface area with the use of Equation 1:

$$A_{appr} \cong px \times GSD \quad (1)$$

where $A_{appr}$ is the surface area of the detected planes and $GSD$ is the ground sample distance. GSD can be calculated using Equation 2:

$$GSD = \frac{W_s \times h \times 100}{FL \times W_I} = \frac{12.75\ mm \times 120\ m \times 100}{10.6\ mm \times 4608\ px} = 3.13\ cm/pxD \quad (2)$$

where $W_s$ is the camera sensor width, h is the altitude of the drone, FL is the focal length and $W_I$ is the width of the camera footprint.

A wide range of airplane producers do not share their surface area values of their planes. Instead, they apply other dimensions that may help in calculating the total surface area. Airports may perform this task and calculate the surface areas for each airplane based on their database. The surface area can finally be used as the key feature for identifying the airplane type. Airplane type identification can be based on the cabin length. The cabin length is shared as a dimension specification by all manufacturers, therefore, it is more suitable to consider as a key feature for identifying each detected airplane. Identifying using the length-based approach can be derived from the surface area-based approach by finding the shortest pairwise distances between pixels within the masked area, then determining the farthest distance. The farthest distance can be computed between pixels located as vertices in a convex hull; thus, using the convex hull algorithm reduces computations to a fewer number of pixels. The convex hull is the smallest polygon surrounding any set of data points on a plane. The polygon is a region of a plane bounded by line segments, called edges, joined end to end in a cyclic manner. The vertices are the points where two successive edges meet. The simplest way to do this is by starting from the leftmost point, which is the point that has the minimum x-coordinate value, then wrapping points in a counter-clockwise direction until reaching the same point.

Fortunately, the collected dataset used in this paper has a list of aircraft types that exist in the images. This list was given by the International Paris Air Show. The aim of this work is to identify the detected airplane type. The evaluation can be performed by comparing the lengths of the output with the real aircraft lengths. The output lengths are calculated by averaging the detected length over the original validation set, as shown in Equation 3:

$$Detected\ length\ avg = \frac{\sum_{j=1}^{j=n} x_j}{n} \quad (3)$$

where j is the number of validation images, x is the validation image and n is the total number of validation sets used in the evaluation. Table 4 lists the abbreviated codes of airplane types and their actual lengths.

TABLE IV. AIRPLANE NAMES, TYPE CODES AND ACTUAL LENGTH

| Plane Full Name | Plane Type Code | Actual Length (m) |
|---|---|---|
| Lockheed-Martin-LM100J | LM100J | 35 |
| GULFSTREAM-G-280 | G-280 | 20 |
| GULFSTREAM-G-550 | G-550 | 29 |
| GULFSTREAM-G-650 | G-650 | 30 |
| Cessna-Citation CJ4 | CJ4 | 16 |
| Cessna-Citation M2 | CM2 | 13 |
| Boeing 787-8 | Bo787 | 57 |
| Airbus A-380 | A-380 | 73 |
| Airbus A-320 | A-320 | 38 |

## IV. RESULTS AND DISCUSSION

An evaluation process is used in this paper to measure the performance of the detection approach. Several points must be clarified before discussing the evaluation steps. In object detection, many metrics exist to evaluate each approach. Most of these metrics had been embedded in common competitions such as PASCAL VOC, COCO and Open Image challenges. These metrics use the mean average position and average recall as the fundamental metrics. The confidence score and Intersection over Union (IoU) are fundamental concepts in the evaluation process. The confidence score is the classifier probability of finding an object inside an anchor box. These two terms are used to determine whether detection is a true positive (TP), a false positive (FP), a true negative (TN) or a false negative (FN). In this paper, COCO metrics have been used for evaluation purposes. COCO has multiple mAP metrics defined by multiple IoU-thresholds. In total, Model 6 performed the best from among the other models. This was expected since Model 6 has a lower learning rate. Model 5 presented perfect results since the maximum proposal regions decreased to 100. Tables 5 and 6 offer promising values based on COCO metrics in comparison to other research outcomes that had implemented satellite images. Table 7 is the benchmark that displays the detection accuracy in satellite images for all eight models. Around 200 images were randomly selected from the NWPU-RESISC45 dataset for airplane categorisation. Although satellite images are different from drone images, the accuracy values seem to be similar to other research results, indicating their reliability. The values shown in the tables signify the reliability of the approach used in this paper

TABLE V. TRAINING DATASET EVALUATION

| Model | Metrics |||||||||||| 
|---|---|---|---|---|---|---|---|---|---|---|---|---|
| | 1 | 2 | 3 | 4 | 5 | 6 | 7 | 8 | 9 | 10 | 11 | 12 |
| 1 | 0.899 | **0.99** | 0.98 | 0.848 | 0.917 | 0.931 | 0.369 | 0.927 | 0.927 | 0.882 | 0.947 | 0.952 |
| 2 | 0.898 | **0.99** | 0.98 | 0.819 | 0.918 | **0.978** | 0.399 | 0.918 | 0.918 | 0.849 | 0.943 | **0.985** |
| 3 | 0.269 | 0.577 | 0.205 | 0.053 | 0.311 | 0.688 | 0.197 | 0.348 | 0.389 | 0.157 | 0.448 | 0.755 |
| 4 | 0.892 | **0.99** | **0.99** | 0.875 | 0.903 | 0.903 | 0.389 | 0.924 | 0.924 | **0.908** | 0.931 | 0.925 |
| 5 | 0.92 | **0.99** | **0.99** | **0.863** | 0.943 | 0.961 | 0.403 | 0.940 | 0.941 | 0.88 | **0.963** | 0.977 |
| 6 | **0.921** | **0.99** | 0.983 | 0.875 | 0.937 | 0.969 | 0.402 | **0.942** | **0.943** | 0.901 | 0.958 | 0.982 |
| 7 | 0.902 | **0.99** | **0.99** | 0.582 | 0.897 | 0.914 | 0.434 | 0.926 | 0.928 | 0.580 | 0.924 | 0.939 |
| 8 | 0.816 | 0.965 | 0.928 | 0.737 | 0.896 | 0.935 | 0.412 | 0.840 | 0.842 | 0.772 | 0.922 | 0.957 |

TABLE VI. VALIDATION DATASET EVALUATION

| Model | Metrics |||||||||||| 
|---|---|---|---|---|---|---|---|---|---|---|---|---|
| | 1 | 2 | 3 | 4 | 5 | 6 | 7 | 8 | 9 | 10 | 11 | 12 |
| 1 | 0.568 | 0.945 | 0.614 | 0.416 | 0.617 | 0.791 | 0.284 | 0.614 | 0.619 | 0.502 | 0.664 | 0.821 |
| 2 | 0.542 | 0.928 | 0.570 | 0.367 | 0.597 | **0.808** | 0.278 | 0.588 | 0.595 | 0.459 | 0.647 | 0.829 |
| 3 | 0.240 | 0.533 | 0.174 | 0.054 | 0.287 | 0.710 | 0.187 | 0.304 | 0.349 | 0.147 | 0.418 | 0.742 |
| 4 | 0.554 | 0.936 | 0.605 | 0.389 | 0.611 | 0.766 | 0.284 | 0.598 | 0.603 | 0.467 | 0.664 | 0.796 |
| 5 | 0.570 | 0.927 | **0.652** | 0.413 | **0.628** | 0.801 | 0.287 | 0.62 | 0.684 | **0.677** | **0.829** |
| 6 | **0.573** | 0.938 | 0.645 | **0.426** | 0.627 | 0.781 | **0.289** | 0.617 | 0.625 | 0.504 | 0.672 | **0.829** |
| 7 | 0.525 | **0.955** | 0.556 | 0.00 | 0.440 | 0.610 | 0.216 | 0.586 | 0.593 | 0.0 | 0.51 | 0.672 |
| 8 | 0.364 | 0.767 | 0.335 | 0.175 | 0.51 | 0.756 | 0.174 | 0.419 | 0.435 | 0.274 | 0.573 | 0.791 |

TABLE VII. SATELLITE EVALUATION AS A BENCHMARK

| Model | Metrics |||||||||||| 
|---|---|---|---|---|---|---|---|---|---|---|---|---|
| | 1 | 2 | 3 | 4 | 5 | 6 | 7 | 8 | 9 | 10 | 11 | 12 |
| 1 | 0.448 | 0.911 | 0.384 | 0.226 | 0.477 | 0.637 | 0.230 | 0.505 | 0.520 | 0.386 | 0.540 | 0.697 |
| 2 | 0.405 | 0.845 | 0.317 | 0.184 | 0.434 | 0.611 | 0.221 | 0.466 | 0.484 | 0.329 | 0.508 | 0.659 |
| 3 | 0.244 | 0.531 | 0.174 | 0.061 | 0.29 | 0.465 | 0.157 | 0.334 | 0.375 | 0.34 | 0.494 | 0.579 |
| 4 | 0.397 | 0.813 | 0.300 | 0.188 | 0.427 | 0.575 | 0.209 | 0.455 | 0.482 | 0.335 | 0.506 | 0.656 |
| 5 | 0.456 | 0.902 | 0.398 | 0.209 | 0.495 | 0.637 | **0.241** | 0.520 | 0.524 | 0.343 | 0.555 | 0.709 |
| 6 | **0.472** | 0.932 | 0.430 | 0.260 | 0.499 | 0.687 | 0.240 | **0.530** | 0.538 | 0.387 | 0.560 | 0.738 |
| 7 | 0.393 | 0.852 | 0.250 | 0.245 | 0.436 | 0.456 | 0.211 | 0.480 | 0.506 | 0.374 | 0.540 | 0.532 |
| 8 | 0.314 | 0.693 | 0.244 | 0.106 | 0.360 | 0.491 | 0.190 | 0.380 | 0.409 | 0.224 | 0.441 | 0.612 |

According to the statistics provided in Table 8, the experiments reveal that the best estimations for length are from Models 2, 4 and 5 with an average accuracy of 89%. Model 6 has approximately the same average (87%) and a perfect detection performance, as mentioned in the previous section. Model 3 failed in detecting some airplane types, therefore, its results were disregarded. The detected length values have been assigned to the closest actual lengths to determine the airplane type. Abbreviated code names are used to make proper output annotations.

TABLE VIII. LENGTH DETECTION ACCURACY %

| Plane | Models ||||||||
|---|---|---|---|---|---|---|---|---|
| | 1 | 2 | 3 | 4 | 5 | 6 | 7 | 8 |
| LM100J | 98 | 99 | 98 | 98 | 99 | 99 | 98 | 99 |
| G-280 | 91 | 99 | 93 | 98 | 98 | 99 | 91 | 80 |
| G-550 | 78 | 78 | 73 | 75 | 78 | 78 | 71 | 78 |
| G-650 | 86 | 95 | 92 | 91 | 91 | 90 | 86 | 98 |
| CJ4 | 91 | 97 | - | 96 | 91 | 91 | 85 | 77 |
| CM2 | 77 | 85 | - | 85 | 92 | 73 | 96 | 69 |
| Bo787 | 97 | 99 | 99 | 99 | 95 | 99 | 60 | 96 |
| A-380 | 99 | 98 | 99 | 100 | 99 | 100 | 88 | 95 |
| A-320 | 53 | 53 | - | 56 | 56 | 56 | 43 | 53 |
| Average | 86 | **89** | - | **89** | **89** | 87 | 80 | 83 |

The outcomes have been evaluated by calculating the accuracy of the detected arithmetic mean versus actual values. The accuracy values displayed promising results. Evaluation through the confusion matrix, also known as the error matrix, also proves the method's performance. It can be seen in Table 9 that some errors in airplane type identification do exist. These errors are associated with airplanes that have similar lengths, for instance, G-550 and G-650. The difference in their length is only by 1 meter. Another reason is that some images contain cropped airplanes which affect accurate length detection. Therefore, properly choosing the place and time when acquiring images is extremely important. This can be solved by scheduling drone flights at specific times and at designated locations. In future, additional detection for wing surface area will be provided to enhance the identification of airplane type.

TABLE IX. CONFUSION MATRIX

| Plane | LM100J | G-280 | G-550 | G-650 | CJ4 | CM2 | Bo787 | A-380 | A-380 |
|---|---|---|---|---|---|---|---|---|---|
| LM100J | 9 | | | 2 | | | | | 1 |
| G-280 | | 20 | 4 | | 2 | | | | 2 |
| G-550 | | 1 | 10 | 3 | | | | | 1 |
| G-650 | 2 | | 2 | 8 | | | | | 1 |
| CJ4 | | 3 | | | 11 | 3 | | | |
| CM2 | | | | | | 31 | | | |
| Bo787 | | | | | | | 12 | 2 | |
| A-380 | | | | | | | | 6 | |
| A-320 | 2 | | | | | | 1 | 2 | 4 |

## V. CONCLUSION

In this paper, an easy and low-cost approach was highlighted to support airplane traffic control at airports. The suggested approach is to use a deep learning model with aerial images collected by drones to detect airplanes as the first step. This approach can be a general solution that can be applied in any country around the world. It only requires drones to feed the system with aerial images rather than using the satellite approach, which requires advanced and expensive technology. Evaluation has yielded promising values in terms of COCO metrics. Although satellite images are different from drone images, the accuracy values seem to be similar, indicating the reliability of this approach. This paper further introduced a reliable airplane type identification using a mask region convolution neural network for airplane detection. It applies surface area and length calculations to the detected object. The length was used as the key factor to determine the type of airplane since this factor was given by the manufacturer as one specification data. This work provides an effective solution for managing airplane traffic jams at airports. Results have shown that the proposed approach is capable of identifying different types of airplanes. Knowing each airplane type at the airport would provide useful information such as physical and technical specifications. Physical specifications can be used to determine the suitable area needed for aircraft allocation. Technical specifications can be used to establish the type of services needed and the time duration. Overall, this work considerably contributes to solving the issue of airport traffic jams using appropriate, simple and low-cost techniques.